\documentstyle[11pt,aaspp4]{article} 
\newcommand{\kms}{km~s$^{-1}$}
\newcommand{\kmsM} {km~s$^{-1}$~Mpc$^{-1}$}
\newcommand{\subsun}{\mbox{$_{\odot}$}}
\newcommand{\etal}{{\it et al.\/}}
\begin{document}

\title{Lost and Found: The Damped Lyman Alpha Absorbers in the QSO OI 363
\altaffilmark{1} }

\author{Judith G. Cohen\altaffilmark{2} }

\begin{abstract}
The galaxy giving rise to the damped Ly$\alpha$ absorbing system in the QSO OI 363
with $z=0.221$ has been found.  A galaxy which is probably associated with
the second DLA in this same QSO at $z=0.0912$ has also been found.  Neither
galaxy is very luminous, and neither galaxy shows signs of extensive current
star formation, a massive disk or lots of gas.  The impact parameters 
for each of the two
galaxies with respect to the QSO are reasonable.  
If most DLA absorbers arise in such low
luminosity galaxies, it will be difficult to pick out the correct
galaxy giving rise to DLA systems at high redshift within the large
projected areal density on the sky of faint galaxies around 
distant QSOs.

\end{abstract}


\section{Introduction}

A long series of investigations reviewd
by Weymann, Carswell \& Smith (1981) (see Bergeron \& Boiss\'e 1991,
Steidel \etal\ 1997,  Churchill \etal\ 1999 for a current update)
has established that when the HI column density is sufficiently large 
($N(HI) > 10^{15}$ atoms cm$^{-2}$),
Ly$\alpha$ absorption seen in the spectra of QSOs is often accompanied by
absorption in the CIV doublet at 1550\AA.  At somewhat higher
column densities ($N(HI) > 10^{17}$ atoms cm$^{-2}$), C II, Si II,
Al II, Fe II, and Mg II are often detected.
Furthermore, the origin of the low-ionization absorbing gas can usually
be identified as a luminous galaxy whose redshift matches that of
the absorbing gas and which has an impact parameter of
$\lesssim150h^{-1}$ kpc with respect to the QSO.
Chen \etal\ (1998) present recent results on identifying
the galaxies that give rise to the lower column density
Ly$\alpha$ forest absorption. 
Such studies are a key way of probing 
gas in the outer regions of galaxies. 

The damped Ly$\alpha$ absorbers (DLAs) are those rare cases with
the highest HI column densities, $N(HI) > 10^{20}$ atoms cm$^{-2}$.
It is thus somewhat puzzling that one of the nearest DLAs,
the system with $z=0.0912$ seen in the spectrum of the QSO OI 363,
remains unidentified.   This DLA was discovered by
Rao \& Turnshek (1998) (henceforth RT) during a large survey 
program to scrutinize HST spectra
of QSOs taken in the UV described in Rao \& Turnshek (2000).  
However, they were unable to identify
the galaxy producing the absorption.  By chance there is a second
DLA in the spectrum of this QSO, at $z=0.2212$, and they were
also unable to establish the origin of this DLA.  The 
HI from these two systems has been detected in absorption
at 21 cm  by
Lane \etal\ (1998) and by Chengalar \& Kanekar (1999).

Rao \& Turnshek (1998) present a deep image of the field of this
QSO, and obtained a spectrum of the brightest galaxy within
30 arcsec of the QSO (galaxy G11,
using their identifications - see figure 4 of their paper), which
shows pronounced spiral arms.  Galaxy G11 turned out to be
a foreground object at $z=0.06$ not associated with
either of the DLA systems. 
The most likely remaining candidate for the DLA, identified
in the deep imaging survey of Le Brun \etal\ (1993),  was too faint
for them to obtain a spectrum.
They state that with the elimination of G11, there is no 
candidate galaxy bright enough to produce these two DLA systems,
and that this indicates an inconsistency
with the standard model developed by Prochaska \& Wolfe (1998)
for DLAs arising in large HI disks
of galaxies.

\section{New Spectroscopy in this Field}

I have 
obtained spectra of the two next brightest candidates within 30 arcsec
of the QSO with the hope of identifying
the absorbing galaxy in each of the two low-$z$ DLA systems seen
in the QSO OI 363.  The properties of the four brightest galaxies
near this QSO are listed in Table 1.  A more complete census of
the galaxies near OI 363 can be found in RT.

I have established that the $z=0.2212$ DLA system originates in 
galaxy G1 which is only 6 arcsec from
the QSO.   This is the galaxy suggested as the possible DLA host
by Le Brun \etal\ (1993). The measured
redshift of this galaxy from a 1500 sec exposure taken on March 6, 2000 with 
the Low Resolution Imaging Spectrograph (Oke \etal\ 1995)
at the Keck Observatory using a 1.5 arcsec slit with a 300 g/mm grating
(spectral resolution 15\AA) is $z=0.221$.  The luminosity of G1 is 
$M_R = -19.3$ (1/7 $L^{\ast}_R$). \footnote{We adopt $H_0 = 70$ \kmsM\
and $\Omega_M = 0.3$}.
The impact parameter with respect to the QSO is 13$h^{-1}$ kpc.

The spectrum of G1 is shown in Figure~1.  Absorption 
in the H+K doublet of CaII is strong, and
a pronounced 4000\AA\ break is seen, as well as the G band of CH
and the ultraviolet CN band.
No emission lines were detected, with an upper limit to
the equivalent width for 
the 5007 \AA\ [OIII] line of 5 \AA, and that for H$\alpha$ of 10 \AA.
The SNR in the continuum per spectral resolution element is 45 
at about 4885 \AA\ (rest frame 3990 \AA) and $\sim$37 at
rest frame 3790\AA, i.e. just above and below the H+K doublet.

The origin of the absorption for the $z=0.0912$ DLA system is
still unclear.
Galaxy G10 is probably associated with this gas in some way, but is
not directly the source.  This galaxy, which RT describe as
an ``early type galaxy'', is 28 arcsec from the QSO.
The color of G10 is somewhat bluer in $B-R$ than that of G11 
(Monet \etal\ 1999).  
The redshift 
of G10 is $z=0.106$. With this redshift, G10 has an impact
parameter of $\sim35h^{-1}$ kpc and a luminosity of
$M_R = -18.2$ (1/14  $L^{\ast}_R$).  Its
spectrum  (a 1000 sec exposure with the
same instrumental configuration as for G1) is also shown in Figure~1.
It appears similar to that of G1, but the 3968 \AA\ line is stronger
relative to that at 3933\AA\ than in galaxy G1.  
The G band of CH appears to be present 
in the spectrum of G10.  
The [OII] emission line at 3727\AA\
is beyond the blue end of the spectrum; the range is 4200 to 8900 \AA.
The SNR here in the continuum per spectral resolution element is between 35 and 40 at wavelengths just below and just above the H+K doublet of CaII.
There is
a possible detection of H$\alpha$ with an equivalent width of
$\sim$3 \AA. The velocity difference between G10
and the DLA itself is 
uncomfortably large ($\sim$4000 \kms) and G10 itself
cannot be the origin of the DLA absorption. 

Thus
the three brightest galaxies within 30 arcsec of the
QSO, G1, G10 and G11, have now been observed spectroscopically,
and none appear to be the source for this DLA gas.
The next brightest galaxy, assuming that no mistake was
made by RT in separating stars from galaxies, within this area
is more than a magnitude fainter than G1, which is the faintest
of the three already observed spectroscopically.
(The field is at a rather low galactic latitutde 
($b^{II} = 23.6^{\circ}$),
so there are many stars in this magnitude range as well.) 

Perhaps G10 is a member of a cluster,
and one of the other galaxies in that group is the actual host for
the absorbing gas.    Hopefully planned future
observations of some of the fainter galaxies nearer the QSO
than G10 will reveal the culprit for this DLA shortly, but the
upper limit on its luminosity is now constrained to be 
(1/30 $L^{\ast}_R$).

\section{Discussion}

As emphasized by
Steidel, Dickinson \& Persson (1994), all normal field galaxies can give
rise to QSO absorption lines, albeit generally
of lower total column density than those
characteristic of DLAs.  Steidel \etal\ (1997) find that
in the regime $z \sim 0.5$ any
galaxy over a wide range of morphological types,
from late spirals to S0s,
with $L > L^{\ast}/10$ can produce MgII absorption and an
associated Ly$\alpha$ forest line.  The HI column
densities required for a DLA system are, however, several orders of
magnitude higher.

Identifications are available for a few other low-$z$ DLAs.
Lanzetta \etal\ (1997) have established that the galaxy responsible for
the $z=0.1638$ absorption in the spectrum of QSO 0850+4400 is
a moderate luminosity ($L^*_B/2.3$) S0 galaxy.  However, the HI
column density for this system (log$N(HI) = 19.81$)
is just below the cutoff normally adopted for DLAs. 
The closest known
DLA, studied by Miller, Knezak \& Bregman (1999),
arises in the outer part of NGC 4203, with Ton 1480
as the background QSO.  NGC 4203 is an isolated E3 galaxy  (de Vaucouleurs, 
de Vaucouleurs \& Corwin 1976) with
$M_B \approx -19.2$, which corresponds roughly to
$1/3 L^{\ast}_B$.

Even though the sample is small, 
it is clear that gas rich galaxies with extensive current star
formation and with luminosities $L > L^{\ast}/3$
are not the dominant origins of the low-$z$ DLA 
systems.  Galaxies of moderate luminosity and without obvious
signs of high gas content from their optical morphology or spectra
appear to give rise to DLA systems when there is a 
background QSO at a suitable impact parameter.

A possible origin for DLA gas is low luminosity, 
gas rich dwarf galaxies such as the Local Group member
NGC 6822, which has a large HI halo
with a mass of about 1.5${\times}10^8$ M\subsun\ in HI (Roberts 1972)
in spite of its low luminosity of $M_V \sim -16.0$.  However, the 
admittedly small sample of known
low-$z$ DLA  galaxies are not gas rich dwarfs.

At intermediate redshift, many DLA candidates are identified from HST imaging
surveys, with few spectroscopic confirmations. Le Brun \etal\ (1997)
and Steidel \etal\ (1995) suggest that the most probable DLA
candidates are galaxies of moderate luminosity.  The latter group
finds that the probable identification for the
DLA with $z=0.8596$ along the line of sight to
PKS 0454+0356 has $M_B = -18.7$, corresponding to
$L^*_B/4$, ignoring any evolution in luminosity of $L^*_B$ with redshift.

At high redshift, 
the suggestion that DLAs arise in  dwarf galaxies has been
made by York \etal\ (1986), 
and more recently by Matteucci, Molaro \& Vladilo (1997), while
Jimenez, Bowen \& Matteuci (1999) have suggested
low surface brightness galaxies as the culprits.
On the other hand, McDonald \& Miralda-Escude (1999) and Haehnelt, Steinmetz
\& Rauch (1998) view the DLAs at high redshift as resulting
from the formation of protogalaxies and ascribe the velocity
widths to turbulence, rather than ordered motions in normal
rotating galactic disks.  Ground baesd long slit
spectroscopy at rest-frame H$\alpha$ and HST imaging (Bunker \etal\ 1999 and Kulkarni \etal\ 2000, respectively)
demonstrate that suspected DLA absorbers at high
redshift appear to have small sizes and low star formation rates.

Thus the model of Prochaska \& Wolfe (1998) of
massive rotating disks does not correspond to the observed properties
of most of the galaxies that give rise to DLA systems at 
low redshift, and this may also hold true at high redshift.

The fact that the
identifications for low-$z$ DLA systems seem to be low luminosity galaxies
which are not starbursts or other very gas rich systems is rather unexpected.
It implies that searches for the galaxies producing DLA systems 
at high redshift will be quite
difficult, as unveiling the right galaxy from the projection
through a long line of sight of the galaxy luminosity
function will not be easy.  Furthermore, since the majority of
field galaxies occur in
groups out to at least $z \sim 1.1$ (Cohen \etal\ 2000), 
one may misidentify the 
distant DLA gas with a galaxy which is
in fact not the true source of the gas, but some brighter member of
the same group or cluster.

Perhaps the selection effects attributed  to dust
discussed by Ostriker \& Heisler (1984) and applied to DLA systems
in particular by Fall \& Pei (1993) and most recently by
Boiss\'e \etal\ (1998) 
are responsible.  They suggest that the higher mean extinction
within massive galaxies with high dust content
and high metallicity act to limit our ability to detect the
background QSOs.  Since dust absorption is higher in the UV,
another consequence of dust would be a 
limitation in our ability
to obtain UV spectra of the background QSOs to 
identify the absorption line systems, an effect dependent on 
$z$(QSO) -- $z$(DLA). 
Such a selection effect would skew the probability
that a galaxy produces a detectable DLA system away
from one based only
on the galaxy's gas column density and impact parameter to the QSO. 
Another consequence of dust might be to bias
calculations based on QSO absorption features
of the contribution of the neutral gas to the cosmological mass
density.  If the next few identifications of low-$z$ DLA systems
continue to be low luminosity galaxies, these selection
effects due to dust may be playing an important role.


\acknowledgements The entire Keck/LRIS user community owes a huge debt
to Jerry Nelson, Gerry Smith, Bev Oke, and many other people who have
worked to make the Keck Telescope and LRIS a reality.  We are grateful
to the W. M. Keck Foundation, and particularly its late president,
Howard Keck, for the vision to fund the construction of the W. M. Keck
Observatory.  I thank C.Steidel and the 
anonymous referee for helpful comments.

\clearpage

\begin{figure}
\epsscale{1.0}
\plotone{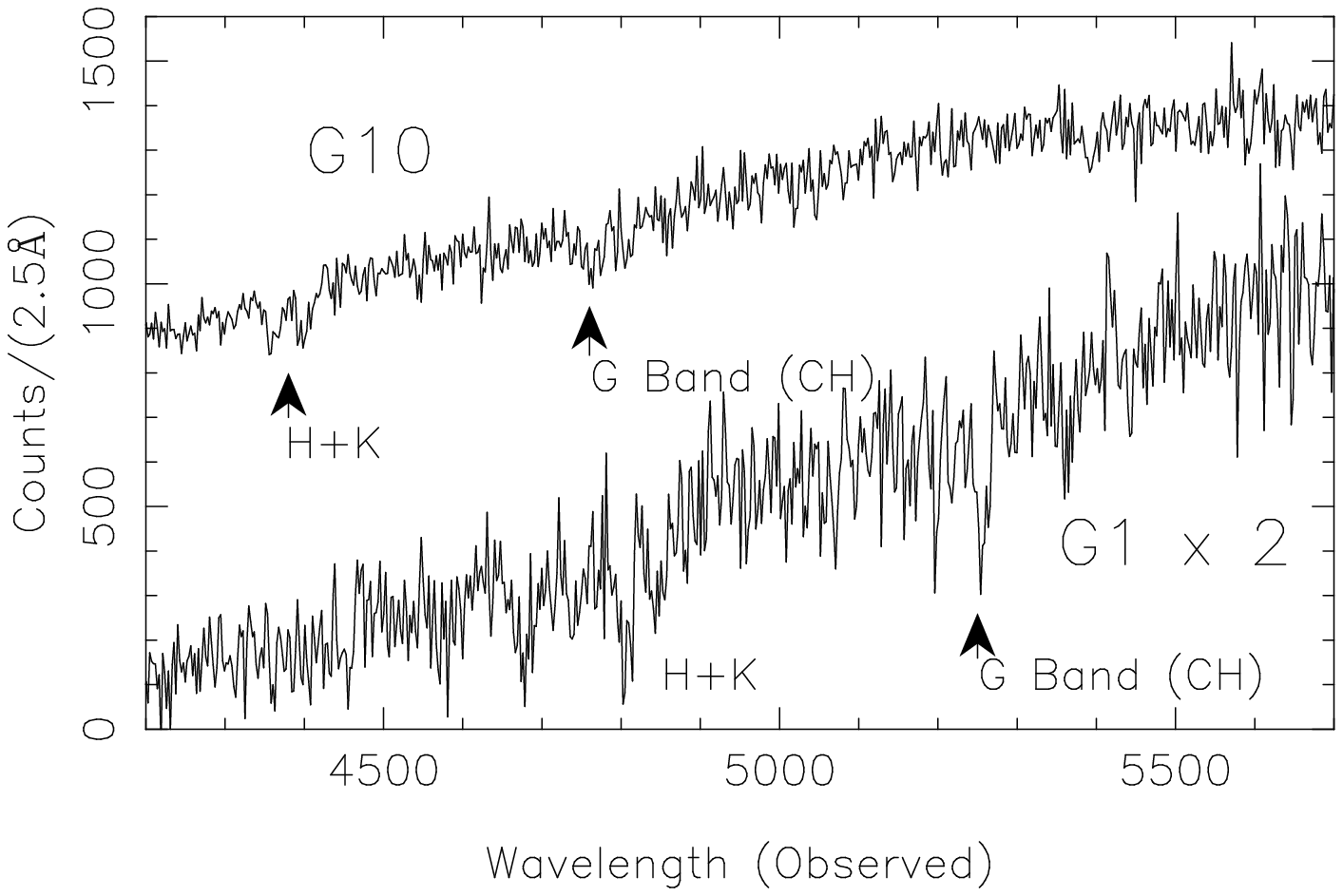}
\caption[figure1.ps]{The LRIS spectra of galaxies G1 and G10
near the QSO OI 363 are shown.  The vertical axis is counts/pixel,
with 2 detected electrons producing 1 count.  The spectrum of galaxy
G1 has been multipled by a factor of 2.0 and there is a vertical
offset of 800 counts between the two spectra.
\label{fig1}}
\end{figure}

\clearpage

%
%
\begin{deluxetable}{lllrr}
\tablenum{1}
\tablewidth{0pt}
\tablecaption{Properties of the Four Brightest Galaxies Near the QSO OI 363}
\label{tab1}
\tablehead{\colhead{ID} & \colhead{$R$\tablenotemark{a}} & \colhead{$z$} & 
\colhead{$M_R$} & \colhead{$\Delta(\theta)$\tablenotemark{a}}  \nl
\colhead{} & \colhead{(mag)} & \colhead{} & \colhead{(mag)} &
\colhead{(arcsec)} \nl
}
\startdata
G11 & 17.1 & 0.06\tablenotemark{a} & $-$20.3 & 31 \nl
G10 & 19.8 & 0.106 &  $-$18.9 & 28 \nl
G1 & 20.8 & 0.221 & $-$19.6 & 6 \nl
G6 & 22.0 & ... & ... & 16 \nl
\enddata
\tablenotetext{a}{From Rao \& Tunshek (1998)}
\end{deluxetable}

\end{document}